# Analysing the Competency of Mathematical Modelling in Physics

**Edward F. Redish**

*Department of Physics, University of Maryland, College Park, MD, USA*

## Abstract

A primary goal of physics is to create mathematical models that allow both predictions and explanations of physical phenomena. We weave maths extensively into our physics instruction beginning in high school, and the level and complexity of the maths we draw on grows as our students progress through a physics curriculum. Despite much research on the learning of both physics and math, the problem of how to successfully teach most of our students to use maths in physics effectively remains unsolved. A fundamental issue is that in physics, we don't just *use* maths, we think about the physical world with it. As a result, we make meaning with mathematical symbology in a different way than mathematicians do. In this talk we analyze how developing the competency of mathematical modeling is more than just "learning to do math" but requires learning to blend physical meaning into mathematical representations and use that physical meaning in solving problems. Examples are drawn from across the curriculum.

## Keywords

Physics education research, mathematics in science, making meaning with mathematics.

## Mathematics: A critical competency for learning physics

Mathematics plays a significant role in physics instruction, even in introductory classes, but not always in a way that is successful for all students. As physics students learn the culture of physics and grow from novice to expert, many have trouble bridging what they learn in math with how we use mathematics in physics. As instructors, many of us are distressed and confused when our students succeed in maths classes but fail to use those same tools effectively in physics. Part of the difficulty is that in physics, we don't just calculate with maths, we "make meaning" with it, think with it, and use it to create new physics. Mathematics has been identified as a critical scientific competency both by the European Union (EUR-LEX 2006) and the US biology community (National Research Council 2003, AAMC/HHMI 2009, AAAS 2011), so as we think about how we might improve physics instruction it is important to try to understand what role maths play in physics, how that role may be difficult for students, and how we might learn to think about that difficulty. A crucial element is the role that mathematics plays in the epistemology of physics.

The process of science and the development of scientific thinking is all about epistemology – deciding what we know and how we decide that we know it. In physics, mathematics has been closely tied with our epistemology for 300 years, transforming physics from natural philosophy into the mathematical science it is today. For those of us who practice physics, either as teach-ers or researchers, our knowledge of physics, what we know and what we believe is true is deeply blended with mathematics. This tie is so tight that we may find it hard to unpack our blended knowledge and understand what it is that students find difficult.

My research group has been studying students using math in physics at the university level for more than 20 years in a number of different contexts:

- Engineering students in introductory physics
- Physics majors in advanced classes
- Biology students in introductory classes, both with mixed populations and in a specially designed class for biology majors and pre-health-care students.

Since we have been trying to develop insight into what is going on in our students' minds, our data is mostly qualitative. It often involves videos of problem-solving interviews or ethnographic data of students in real classes solving real homework problems, either alone or in groups. Sometimes we have quantitative data as well, including responses of many students on multiple-choice questions on exams or with clickers in a large lecture class.

We work in the theoretical framework of *Resources* – the idea that student thinking is highly dynamic, calling on multiple smaller bits of knowledge that may be organized in locally coherent, but often changing ways. (Hammer 2000) This framework is built on ideas from education, psychology, neuroscience, sociology, and linguistics research. (Redish 2014)

## Different Languages: Math in physics is not the same as math in math

We often say that "mathematics is the language of physics", but what physicists do with maths is deeply different from what mathematicians do with it. Mathematicians and physicists load meaning onto symbols differently and this has profound implications. (Redish & Kuo 2015).

In physics, we link our equations to physical systems and this adds information on how to interpret them. Our symbols carry extra information not present in the abstract mathematical structure of the equation. As a result, our processing of equations in physics has additional levels and may be more complex than the processing of similar equations in a math class.

In physics most of our symbols don't stand for numbers (or collections of numbers) but for measurements. Our symbols bring physical properties along with them. As a result, they have *units* that depend on the measurement process. In math terms, this is quite sophisticated. As a result of the arbitrariness in our choice of units, physics equations must have a particular structure. Since the choice of scale is arbitrary, any physically true relation must be true whatever choice of scale is made. This means that every part of both sides of the equation must change in the same way when a scale is changed. Mathematically this means that equations must transform properly – covariantly by an irreducible representation of the 3-parameter scaling group SxSxS for units of mass, length, and time. (Bridgman 1922)

What about significant figures? Why do we bother talking about them now that we have calculators? But when we multiply 5.42 x 8.73 in a 6th grade arithmetic class we want something different from what we want when we are measuring the area of a (5.42 cm) x (8.73 cm) sheet of silicon. Every physical measurement has an uncertainty that propagates to the product, leaving many digits shown by a calculator as "insignificant figures", irrelevant to physical silicon.

An elementary example of how the physicist's mapping of physical meaning onto symbols changes the way equations are interpreted is illustrated in the example shown in Figure 1. This problem was given as a clicker question to a class of about 200 students in algebra-based introductory physics as part of a lecture on the electric field. The topic had been discussed in a pre-vious lecture and the students had been presented with a derivation of the electric field from Coulomb's law for the electric force of many source charges acting on a test charge.

They had seen the equations

$$\vec{F}_{q_0} = \frac{k_C q_0 q_1}{r_{01}^2} \hat{r}_{1\to0} + \frac{k_C q_0 q_2}{r_{02}^2} \hat{r}_{2\to0} + \frac{k_C q_0 q_3}{r_{03}^2} \hat{r}_{3\to0} \cdots + \frac{k_C q_0 q_N}{r_{0N}^2} \hat{r}_{N\to0}$$

$$\vec{E}(\vec{r}) = \frac{\vec{F}_{q_0}}{q_0}$$

and had pointed out to them that the charge of the test charge factors out and cancels when defining the electric field. When asked, most students could cite the result, "The electric field is independent of the test charge that measures it."

A very small charge $q_0$ is placed at a point $\vec{r}$ somewhere in space. Hidden in the region are a number of electrical charges. The placing of the charge $q_0$ does not result in any change in the position of the hidden charges. The charge $q_0$ feels a force, $F$. We conclude that there is an electric field at the point $\vec{r}$ that has the value $E = F/q_0$.

If the charge $q_0$ were replaced by a charge $-3q_0$, then the electric field at the point $\vec{r}$ would be

      a) Equal to $-E$

      b) Equal to $E$

      c) Equal to $-E/3$

      d) Equal to $E/3$

      e) Equal to some other value not given here.

      f) Cannot be determined from the information given.

**Figure 1.** A quiz problem that students often misinterpret.

Nonetheless, when asked the question in Figure 1, nearly half chose answer (c). These students treated the physics as a pure math problem: "If A = B/C what happens to A if C is replaced by −3C?" They ignored the fact that $F$ here is not a fixed constant, but represents the force felt by charge $q_0$ and therefore implicitly depends on the value of $q_0$.

A second example illustrates another of the differences between maths and physics classes. Maths classes typically use equations with a small number of symbols, with fixed conventions for what symbols stand for variables and what for constants. Furthermore, introductory maths classes (through calculus) often do very little with parameter dependence. In physics, on the other hand, our equations often involve a blizzard of symbols, some of which may be variables or constants depending on what problem we choose to consider. An example occurred in an introductory physics class for life scientists. One year of calculus was a pre-requisite and most of the students in the class had earned a good grade in that class. Nonetheless, many had trouble knowing how to approach the problem shown in Figure 2. The problem was presented as a "work together" problem in a large lecture in which I was serving as a facilitator. As I walked around the class, watching and listening to students, I found one group totally stuck.

When a small organism is moving through a fluid, it experiences both viscous and inertial drag. The viscous drag is proportional to the speed and the inertial drag to the square of the speed. For small spherical objects, the magnitudes of these two forces are given by the following equations:

$$F_v = 6\pi\mu R v$$

$$F_i = C\rho R^2 v^2$$

For an organism (of radius $R$) is there ever a speed for which these two forces have the same magnitude?

**Figure 2.** An example of multiple-parameter use in a physics class for life science students.

**Student:** I don't know how to start. Should I see if I can find all the numbers on the web?

**Facilitator:** *Well, it says 'Do they ever have the same magnitude?' How do you think you ought to start?*

**S:** Set them equal?

**F:** *OK. Do it.*

**S:** I don't know what all these symbols mean.

**F:** *Well everything except the velocity are constants for a particular object in a particular situation.*

**S:** …..[concentrating for almost a minute...] Oh! So if I write it .... $Av = Bv^2$... Wow! Then it's easy!

I have seen many introductory students having serious trouble with the multi-parameter equations in physics and have seen that these same students can easily carry out analogous mathematical problems with only variables and numbers.

## Making Meaning with Mathematics

Our examples suggest that the critical difference in maths-as-pure-mathematics and maths-in-a-physics-context is the blending of physical and mathematical knowledge. A simple model (Redish 2005) focuses on a few of the main steps: (1) choosing a model to map physical quantities into mathematical structures, (2) processing, using the tools inherited from those mathematical structures, (3) interpreting the results back in the physical world, and (4) evaluating whether the result is adequate or whether the original model needs to be refined.

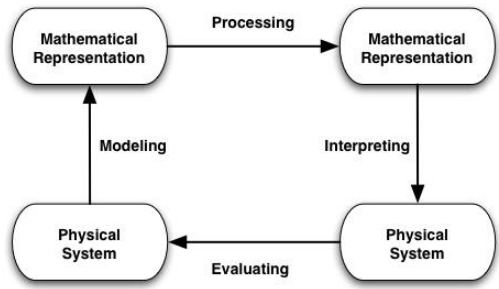

**Figure 3.** A model of mathematical modeling

Often these all happen at once – are intertwined. (The diagram is not meant to imply a step-by-step algorithmic process). In physics classes, processing is often stressed and the remaining elements short-changed or ignored. But in physics, maths integrates with our physics knowledge and does work for us. It lets us carry out chains of reasoning that are longer than we can do in our head, by using formal and logical reasoning represented symbolically. Some of the things we do with maths include

- Calculations
- Predictions
- Summary and description of data
- Development of theorems and laws

But maths in physics also codes for conceptual knowledge, something that is typically not part of what is learned in a mathematics class, such as:

- Functional dependence
- Packing concepts
- Epistemology

An example of how we use equations to organize and pack our conceptual knowledge is shown in Figure 4: Newton's second law. When we just write "$F=ma$", our students may see it simply as a way to calculate either $F$ or $a$ and miss the deeper meaning.

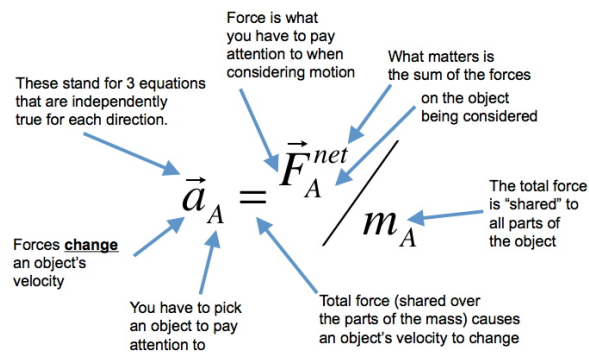

**Figure 4.** Conceptual knowledge packed in an equation. (From NEXUS/Physics.)

## What Does "Meaning" Mean? Some advice from cognitive science

In physics, we "make physical meaning" with maths. Mathematics is a critical piece of how we decide we know something - our epistemology. What does that mean and how does it work?

To develop an answer, we first ought to consider the question, "How do we make meaning with words?" We'll draw on cognitive semantics – the study of the meaning of words in the intersection of cognitive science and linguistics. Some key ideas developed in these fields are relevant:

- *Embodied cognition* – Meaning is grounded in physical experience. (Lakoff & Johnson, 1980/2003)

- *Encyclopaedic knowledge* – Webs of associations build meaning. (Langacker 1987)

- *Contextualization* – Meaning is constructed dynamically in response to perceived context. (Evans & Green 2006)

- *Blending* – New knowledge can be created by combining and integrating distinct mental spaces. (Fauconnier & Turner 2003)

One way embodiment allows maths to feel meaningful in maths is with symbolic forms (Sherin 2001, Redish & Kuo 2014): associating symbol structure with relations abstracted from (embodied) physical experience

- Parts of a whole:  $\square = \square + \square + \square$ ...
- Base + change:  $\square = \square + \triangle$
- Balancing:  $\square = \square$

A second way maths build meaning is through association via multiple representations

- Equations
- Numbers
- Graphs

Physicists tend to make additional meaning of mathematical symbology by associating symbols with physical measurements. This allows connections to physical experience and associations to real world knowledge. And that knowledge may be built up as students learn physics.

But just as we saw with introductory students, students at more advanced levels may not apply knowledge they have about the physical world in a math problem. Figure 5 shows an example drawn from an upper division electricity and magnetism class for physics majors. Our data is taken from a video of two students working on a problem from their text (Griffiths 1999).

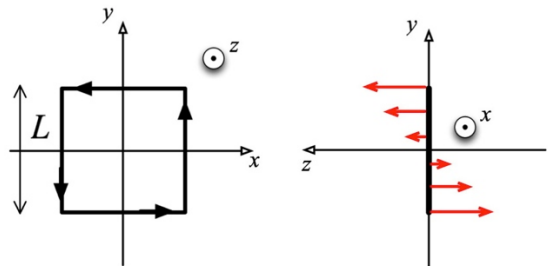

**Figure 5.** An E&M problem. Find the force on a current-carrying loop of wire in a space-varying magnetic field, $\vec{B} = B_0 y \hat{k}$ shown in red. (Griffiths 1999).

Students A and B have independently solved the problem and begin to discuss how they did it. Student A thinks there is a net force, student B does not.

| *Student A* | *Student B* |
| --- | --- |
| Huh! Looks pretty simple – like a physics 1 problem. The sides cancel so I can just do $$\vec{F} = I\vec{L} \times \vec{B}$$ on the top and bottom where B is constant. Both give "up" so I'm gonna get $$\vec{F} = IL^2 B_0 \hat{j}$$ | I'm pretty sure they want us to do the vector line integral around the loop. $$\vec{F} = \oint_{\square} I \, d\vec{L} \times \vec{B}$$ It's pretty straightforward. The sides do cancel, but I get the top and bottom do too, so the answer is zero. |

What do you think happened next?

Student A immediately folded his cards in response to student B's more mathematically sophisticated reason and agreed she must be right. Both students valued (complex) mathematical reasoning (where they could easily make a mistake) over a simple and compelling argument (where it's hard to see how it could be wrong) that blends mathematical and physical reasoning.

The students' expectations that the knowledge in the class was about learning to do complex math was supported by many class activities. They both focused on the "Processing step in the 4-box model," just as the professor had in the lectures.

### Analysing Mathematics as a Way of Knowing: Epistemological resources

We can develop a more nuanced view of what is going on. The example shown in Figure 6 is taken from a homework problem in third year course in the Methods of Mathematical Physics (Bing & Redish 2009).

A rocket is taken from a point A to a point B near a mass m. Consider two (unrealistic) paths 1 and 2 as shown. Calculate the work done by the mass on the rocket on each path. Use the fundamental definition of the work,

$$W_{A \to B} = \int\limits_A^B \vec{F} \cdot d\vec{r}$$

not potential energy. Mathematica may or may not be helpful. Feel free to use it if you choose (though it is not necessary for the calculations required).

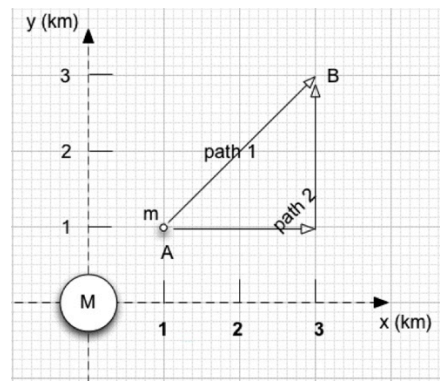

**Figure 6.** A mechanics problem to demonstrate that potential energy is independent of path.

During this discussion three students are talking at cross-purposes. They are each looking for different kinds of "proofs" than the others are offering. They use different kinds of reasons (warrants[1]) to support their arguments. Eventually, they find mutual agreement – after about 15 minutes of discussion.

>**S1:** What's the problem? You should get a different answer from here for this... (Points to each path on diagram)

>**S2:** No no no

>**S1:** They should be equal?

>**S2:** They should be equal

>**S1:** Why should they be equal? This path is longer if you think about it. (Points to two-part path) {**Matching physical intuition with the math**}

>**S2:** Because force, err, because work is path independent. {**Relying on authority – a remembered theorem**}

---

[1] A "warrant" is a specific reason presented to justify a claim. (Toulmin 1958)

**S1:** Well, OK, well is this— what was the answer to this right here? (Points to equation they have written on the board)

$$\int_{\sqrt{2}}^{3\sqrt{2}} \frac{1}{r^2}\,dr = \int_1^3 \frac{1}{y^2+9}\,dy + \int_1^3 \frac{1}{x^2+1}\,dx$$

**S2:** Yeah, solve each integral numerically {**Relying on validity of mathematical calculation**}

**S1:** Yeah, what was that answer...I'll compare it to the number of...OK, the y-one is point one five.

A number of different structures are brought to bear here: different kinds of reasons or *epistemological resources* – such as "I know a theorem" or "This is what the calculation tells us" (Elby & Hammer 2001, Hammer & Elby 2003). We've already seen a number of different ways of coming to a conclusion in a physics problem: the students in Figure 1 who relied on a calculation, those in Figure 5, one who relied on a physical "hands on" (right-hand rule) intuition learned in an earlier physics class, and a second who relied on a complex calculation. The three students solving the problem in Figure 6 first called on different resources – looking at the physical structure of the problem, relying on calculation, and calling up a theorem. Some of the resources commonly we have seen used in a physics class include:

1. *Physical intuition*: Knowledge constructed from experience and perception is trustworthy. (diSessa 1993)

2. *Calculation can be trusted*: Algorithmic computational steps lead to trustable results.

3. *By trusted authority:* Information from an authoritative source can be trusted.

4. *Physical mapping to math*: A mathematical symbolic representation faithfully characterizes some feature of the physical or geometric systems it is intended to represent.

5. *Fundamental laws*: There are powerful principles that can be trusted in large numbers of circumstances (occasionally, all).

6. *Toy models*: Highly simplified examples can yield insight into complex calculations.

Except for the first, these typically involve math, even in an introductory physics class. We also identify a meta-epistemological resource:

7. *Coherence*: Multiple ways of knowing (epistemological resources) applied to the same situation should yield the same result.

**Choosing resources: Epistemological framing**

Our brains know lots of things, and we have many resources for solving our problems, both in life and in a physics class. But the amount of knowledge that can be held in one's mind and manipulated at any instant is limited (Baddeley 1998). The process by which relevant memories and knowledge are brought to the fore is called *framing* (Tannen 1994). When that framing is particularly concerned with knowledge building or problem solving, I refer to it as *epistemological framing* (Hammer et al. 2005, Bing & Redish 2012). Depending on how students interpret the situation they are in and their learned expectations, they may not think to call on epistemological resources they have and are competent with.

Students' epistemological framing can take many forms:

- "I'm not allowed to use a calculator on this exam."
- "It's not appropriate to include diagrams or equations in an essay question."
- "This is a physics class. He can't possibly expect me to know any chemistry."

These are all conscious and easily articulated. Often (as in some examples above), students are not aware of the epistemological choices they have unconsciously made. Epistemological framing can also coordinate significantly with affective responses. (Gupta and Elby 2011)

This epistemological language provides nice classifications of reasoning – both what we are trying to teach and what students actually do. And it can tell us that our assumption that a student failure represents a "student difficulty" with understanding the material may be a misinterpretation of what is going on. There can be epistemological reasons for a student error as well as conceptual ones. But can an epistemological lens provide guidance for instructional design? It can become especially important when students and faculty have different ways of knowing.

## Case Study: Implications for interdisciplinary instruction

Our next example come from NEXUS/Physics (Redish et al. 2014), an introductory physics class developed to meet the needs of biology and life sciences (pre-health care) students. The class is intended to articulate with the rest of these students' curriculum, so calculus, biology, and chemistry are pre-requisites. This allows us to find places where physics has authentic value for biology students – places where they have difficulty making sense of important but complex issues such as chemical bonding (Dreyfus et al. 2014), diffusion (Moore et al. 2014), or entropy and free energy (Geller et al. 2014). The goals of the course are to (1) create prototype open-source instructional materials that can be shared, (2) focus on interdisciplinary coordination of instruction in biology, chemistry, physics, and math, and (3) emphasize competency-based instruction, building general scientific skills. Since physics uses maths heavily, it's an appropriate place in the curriculum to emphasize how maths are used in science.

The course was built with extensive negotiations among all the relevant disciplines (Redish & Cooke 2013, Redish et al. 2014). One of the important things we learned form these negotiations is that the epistemological resources biology students were comfortable using differed from those expected in a physics class. Some resources common in introductory biology are:

1. *Physical intuition*: Knowledge built from experience and perception is trustworthy.

2. *Life is complex*: Living organisms require multiple related processes to maintain life.

3. *Categorization and classification*: Comparison of related organisms yields insight.

4. *By trusted authority:* Information from an authoritative source can be trusted.

5. *Naming is important*: Many distinct components of organisms need to be identified, so learning a large vocabulary is useful.

6. *Heuristics*: There are broad principles that govern multiple situations.

7. *Function implies structure*: The historical fact of natural selection leads to strong structure-function relationships.

In introductory biology, typically none of these involve any math at all. The resources common to the physics list – physical intuition and authority – usually have a mathematical component in introductory physics (e.g., "authority" is often a theorem or equation) while in bio they do not. Even more problematical, two of the critical resources often used in physics, the value of toy models and the power of fundamental laws (mathematically stated), are not only weak or missing in many bio students, they see them as contradicting resources they value – life is complex and function implies structure.

This difference between student and teacher's expectations requires some dramatic changes in our instructional approach from that taken in a traditional physics class. We cannot take for granted that students will value toy models. We have to justify their use. We cannot take for granted that students will understand or appreciate the power of principles (Newton's laws, conservation laws of energy, momentum, or charge). We have to teach not just the content but the epistemology explicitly. We have to create situations in which students learn to see the value of bringing in physics-style thinking with biology-style thinking in order to gain biological insights ("biologically authentic" examples). (Watkins et al. 2012)

**Example: Disciplinary epistemological framing: Why do bilayers form?**

One goal of NEXUS/Physics was to design lessons that explicitly demand a resolution between epistemological resources emphasized in introductory biology and physics. An example of this is a recitation activity on membrane formation. Recitations in this class are done as group work and often require that students bring in their knowledge of biology and chemistry. In this lesson, the question is raised: "Given that the electrostatic attraction between water molecules and a lipid (oil) molecule is stronger than the attraction between two lipid molecules, why does oil and water separate, and, important for biology, how can lipid membranes form?" A videotaped discussion of a student group illustrated not only the mixing of epistemological resources from physics and biology, it shows that the students perceive the mixing of kinds of reasons.

> **S1:** In terms of bio, the reason why it forms a bilayer is because polar molecules need to get from the outside to the inside.
>
> **S2:** If it's hydrophobic and interacting with water, then it's going to create a positive Gibb's free energy, so it won't be spontaneous and that's bad..[proceeds to unpack in terms of positive (energetic) and negative (entropic) contributions to the Gibbs free energy equation.]
>
> **S1:** I wasn't thinking it in terms of physics. And you said it in terms of physics, so it matched with biology.

The first student argues that it has to form because the end result is needed – a typical structure-function argument used in biology. The second student brings in the equation for free energy, $\Delta G = \Delta H - T\Delta S$, presented and analysed both in physics and chemistry, and sees, guided by the structure of the equation, that there is a competition between two effects – energy (here, enthalpy) and entropy. The attraction, being a potential energy, contributes to the enthalpy term, $\Delta H$. But the entropy term involves both lipids and water, and the change of the entropy of the water overcomes the enthalpy term. The students' use of multiple (and interdisciplinary) epistemological resources is illustrated in Figure 7.

**Example: Interdisciplinary instruction or Teaching physics standing on your head**

Both students and faculty may have developed a pattern of choosing particular combinations of epistemological resources in their framing of tasks within a particular class. This leads us to our third epistemological structure: the *epistemological stance*. By this we mean that particular patterns of epistemological framing may become "comfortable" for an individual, with the result that they are likely to frequently activate it as their "normal" or "go-to" knowledge-building approach. The epistemological stances chosen by physics instructors and physics students may be dramatically different – even in the common context of a physics class.

In Figure 8, I show an example that illustrates this. This task was given as a clicker question in a NEXUS/Physics lecture in a discussion meant to extend what the students had learned about potential energy to the case of atom-atom interactions and chemical bonding.

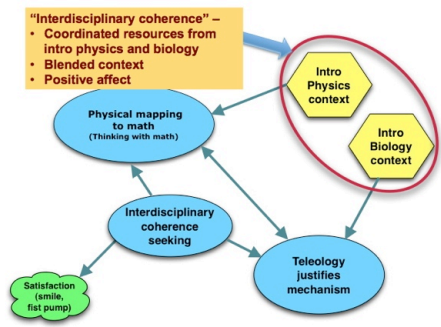

**Figure 7.** The interdisciplinary epistemological resources
used by students in the NEXUS/Physics lesson on membrane formation.

The figure at the left below shows the potential energy of two interacting atoms as a function of their relative separation. If they have the total energy shown by the red line, is the force between the atoms when they are at the separation marked C attractive or repulsive?

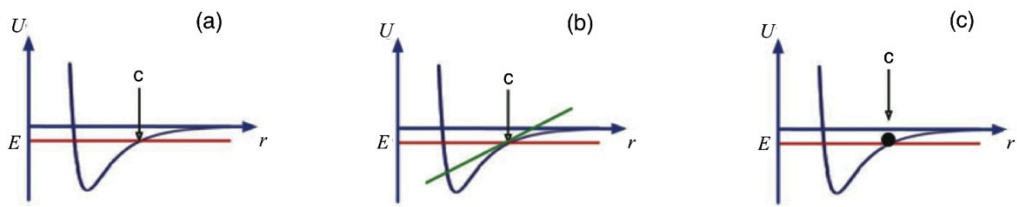

**Figure 8.** Three figures illustrating different epistemological approaches to an explanation.
(a) The figure shown in the problem; (b) an explanation based on a formula;
(c) an explanation based on a physical analogy.

I served as facilitator in two different classes. Two different professors explained it this way when students got stuck: "Remember! $\vec{F} = -\vec{\nabla}U$, or in this case, $F = -dU/dr$. At C, the slope of the $U$ graph is positive. Therefore the force is negative – towards smaller $r$. So the potential represents an attractive force when the atoms are at separation C." They built on the equation that generates Figure 8(b), though both wrote the equation on the board but not the figure.

Wandering around the class while students were considering the problem, I got a good response using a different approach. "Think about it as if it were a ball on a hill [as in Figure 8(c)]. Which way would it roll?  Why? What's the slope at that point? What's the force? How does this relate to the equation $F = -dU/dr$?"

A conflict between the epistemological stances of instructor and student can make teaching more difficult. Physics instructors seem most comfortable beginning with familiar equations – which we use not only to calculate with, but also to remind us of conceptual knowledge. The chain of epistemological resources being used by the professors is something like the following:

*By trusted authority $\Rightarrow$ Calculation can be trusted $\Rightarrow$ Physical mapping to math.*

Most biology students lack experience blending math and conceptual knowledge. They were more comfortable starting with physical intuitions, like, "How does a ball roll on a hill?"

*Physical intuition $\Rightarrow$ Physical mapping to math $\Rightarrow$ Mathematical consistency*

For physicists, math is often the "go to" epistemological resource – the one activated first and the one brought in to support intuitions and results developed in other ways. For biology students, the math is decidedly secondary. Structure/function relationships tend to be the "go to" resource. Part of our goal in teaching physics to second year biologists is to improve their understanding of the potential value of mathematical modelling. This means teaching it rather than assuming it.

## Conclusion

I have presented an analysis of how mathematics is used in physics, including both an unpacking of what professionals do and an analysis of how students respond. I have shown that this can both can give insight into student difficulties reasoning with mathematcis and potentially provide guidance for how to focus on epistemological issues that might create barriers between what a physics instructor is trying to teach and what the students are learning.

We have developed three ways to talk about how students use knowledge, and mathematical knowledge in particular. (1) *Epistemological resources* – Generalized categories of "How do we know?" warrants; (2) *Epistemological framing* – The process of deciding what e-resources are relevant to the current task (NOT necessarily a conscious process); and (3) *Epistemological stances* – A coherent set of e-resources often activated together. But be careful! These are NOT intended to describe distinct mental structures. Rather, we use them to emphasize different aspects of what may be a unitary process: activating a subset of the knowledge you have in a particular situation. A *warrant* focuses on a specific argument, using particular elements of the current context. ("Since the path integral of a conservative force is path independent, these two integrals will have the same value.") A *resource* focuses on the general class of warrant being used. ("You can trust the results in a reliable source such as a textbook.") *Framing* focuses attention on the interaction between cue and response. ("You need to carry out a calculation here.")

Such an analysis has implications for how we understand what our students are doing, what we are actually trying to get them to learn, and (potentially) how to better design our instruction to achieve our goals.

## Acknowledgements

The author gratefully acknowledges conversations and collaborations with the members of the NEXUS/Physics team and the University of Maryland's Physics Education Research Group. This material is based upon work supported by the Howard Hughes Medical Institute and the US National Science Foundation under Awards No. DUE-12-39999 and DUE-15-04366. Any opinions, findings, and conclusions or recommendations expressed in this publication are those of the author and do not necessarily reflect the views of the National Science Foundation.

Edward F. Redish
Department of Physics
University of Maryland
College Park, MD 20742-4111
USA
e-mail: redish@umd.edu